\documentstyle[11pt,newpasp,twoside,epsf]{article}
\def\edcomment#1{\iffalse\marginpar{\raggedright\sl#1\/}\else\relax\fi}
\reversemarginpar

\begin{document}

\newcommand{\chandra}{{\it Chandra} }
\newcommand{\asca}{{\it ASCA} }
\newcommand{\pspc}{{\it Rosat} PSPC }
\newcommand{\xmm}{{\it Newton-XMM} }

\title{Cluster cores as observed with \chandra}
\author{S. Ettori and A.C. Fabian}
\affil{
Institute of Astronomy, Madingley Road, CB3 0HA Cambridge}

\begin{abstract}
We review recent results from \chandra observations of the
X-ray bright cores of clusters of galaxies. We discuss
the detection of ``cold fronts'' and their implication for thermal
conduction, the interaction of cooling flow and radio source
in the Perseus cluster and the dynamical state of the central region 
of A1795. The radial distribution of metals in A1795 and 
the azimuthal variation in metallicity in 4C+55.16 are presented.
\end{abstract}

\section{Introduction}
At radii of few hundred kpc, the intracluster medium (ICM) 
has a characteristic density $n_{\rm gas} \sim 10^{-3}$ cm$^{-3}$, 
temperature $T_{\rm gas}\sim 10^8$~K and a heavy element abundance 
of about 40 per cent of the solar value. 
The density drops at larger radii $r$ approximately as $r^{-2}$. 
Hence, under these conditions, the gas appears as an optically 
thin plasma in ionization equilibrium emitting X-rays, 
where the ionization and emission processes result mainly 
from collisions of ions with electrons. 
This emission is mostly due to thermal bremsstrahlung
(free-free emission)
when $T_{\rm gas} > 3\times 10^7$ K, with an emissivity 
$\propto n_{\rm gas}^2 T_{\rm gas}^{1/2}$.
In the core, where the density is higher, a larger amount of energy 
is radiated away and cooling takes place on timescale
\begin{equation} 
t_{\rm cool} = 1.4 \times 10^8 \ \left( \frac{kT_{\rm gas}}{2 {\rm
keV}} \right)^{0.5} \left( \frac{n_{\rm gas}}{0.1 {\rm cm}^{-3}}
\right)^{-1} \ {\rm yrs} < t_{\rm age} \sim t_{\rm Hubble}. 
\end{equation}
This inequality is satisfied in the core of about 70 per cent 
of nearby clusters (Peres et al. 1998), like A1795 (Fig.~1).
The drops in temperature and the small cooling time suggest
that a subsonic flow of gas moves inward in the hydrostatic core
atmosphere under the influence of gravity and the weight of 
overlying plasma (see Fabian 1994 for a review on this
{\it cooling flow} scenario). 
The inferred deposition rates range from $<$ 10 $M_{\sun}$ yr$^{-1}$
in poor clusters to $>$ 1000 $M_{\sun}$ yr$^{-1}$ in massive
systems. However, only 10 per cent of these cooling rates 
can be accounted for from normal star formation 
(see Crawford et al. 1999 and references therein).

\begin{figure}[hbt]
\plottwo{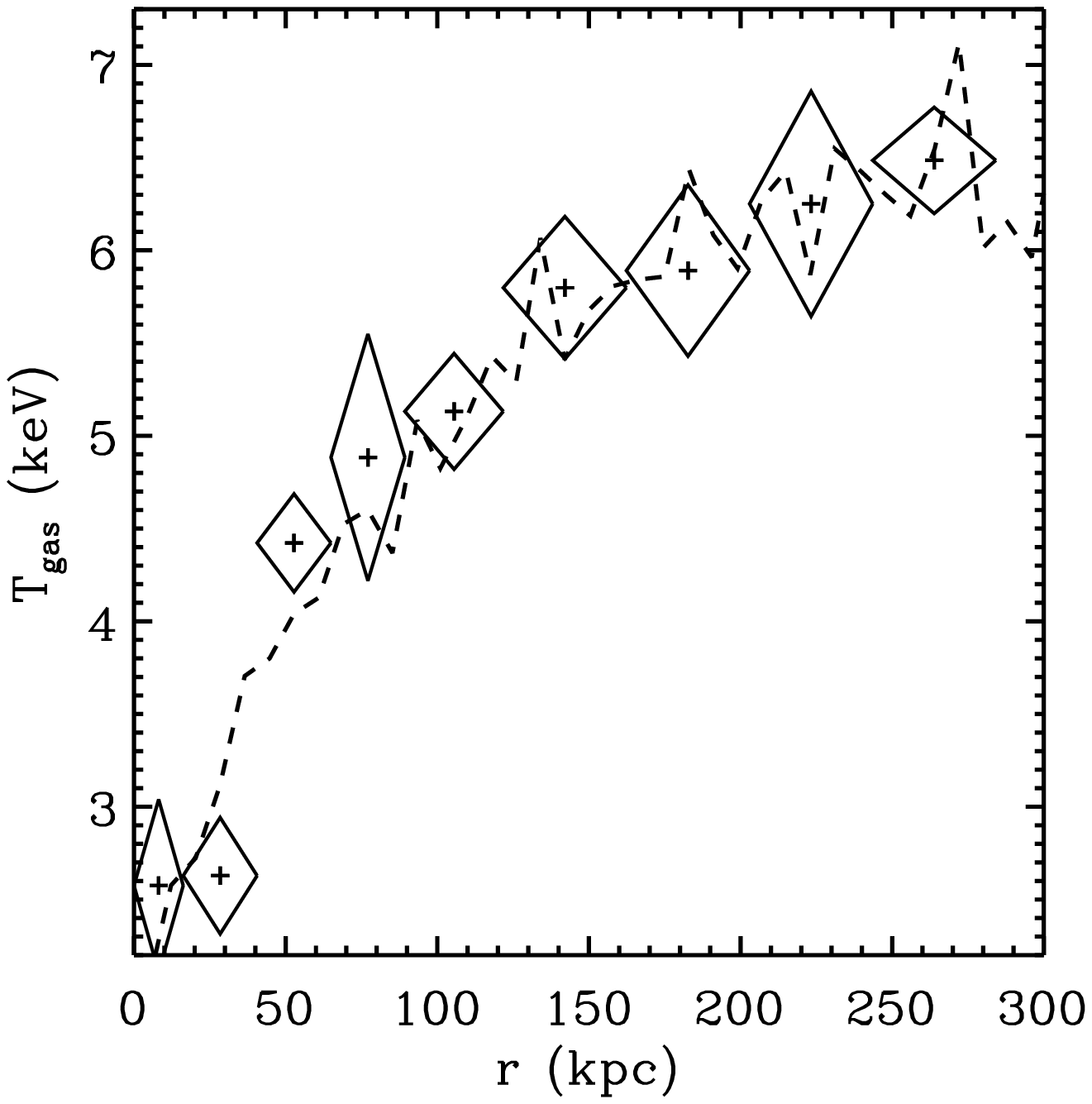}{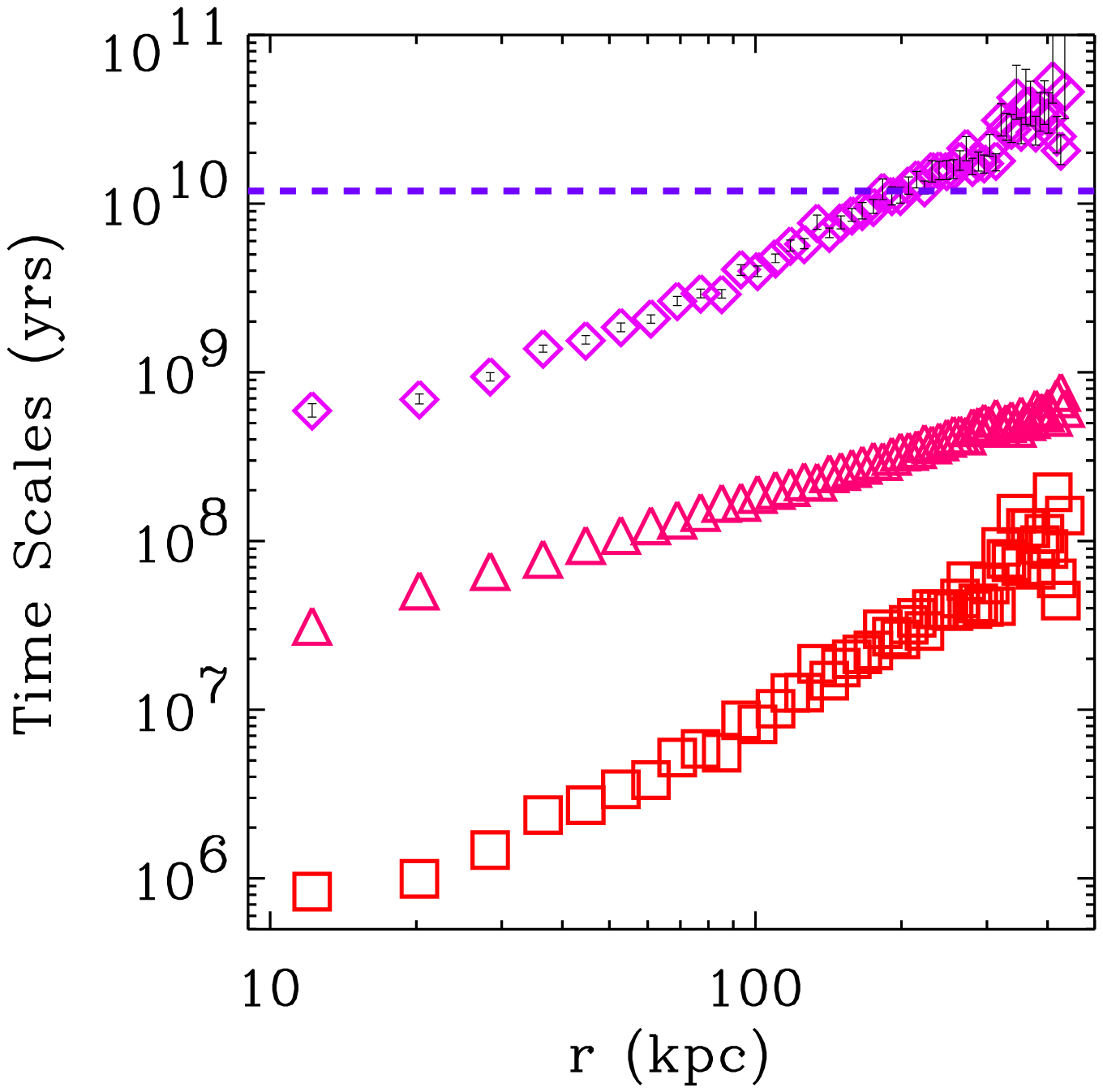}
\caption{\chandra observation of A1795.
(Left) Temperature profile from both 
spatial (i.e. inverting the surface brightness given an assumed
potential law and the hydrostatic equilibrium; {\it dashed line}) 
and spectral (i.e. directly from spectral fitting; {\it diamonds}) 
deprojection analyses (Ettori et al. 2001).
(Right) 
The cooling time, $t_{\rm cool}$ ({\it diamonds}), the
sound crossing time ({\it triangles}) and 
the equipartition time by Coulomb collisions ({\it squares})
are compared to the age of the Universe ({\it dashed line};
$H_0 = 50$ km s$^{-1}$ Mpc$^{-1}$, $\Omega_{\rm m}$=1).
} \label{times} \end{figure}

Here we review some recent results from \chandra observations of 
the cluster central regions on
(i) the physical condition of the plasma and
(ii) how the metals are distributed and
the gas enrichment is produced.

\section{Cold fronts and thermal conduction in cluster cores}

The \chandra observations of ``cold fronts'' in some 
clusters (e.g. A2142, Markevitch et al 2000; A3667,
Vikhlinin, Markevitch, \& Murray  2001; RXJ1720+2638, Mazzotta et al. 2001),
where regions of plasma with differences in temperature by a factor of 2
lie ones close to each other, allow a direct constraint on the
efficiency of the thermal conduction (Ettori \& Fabian 2000).

\begin{figure}[htb]
\plottwo{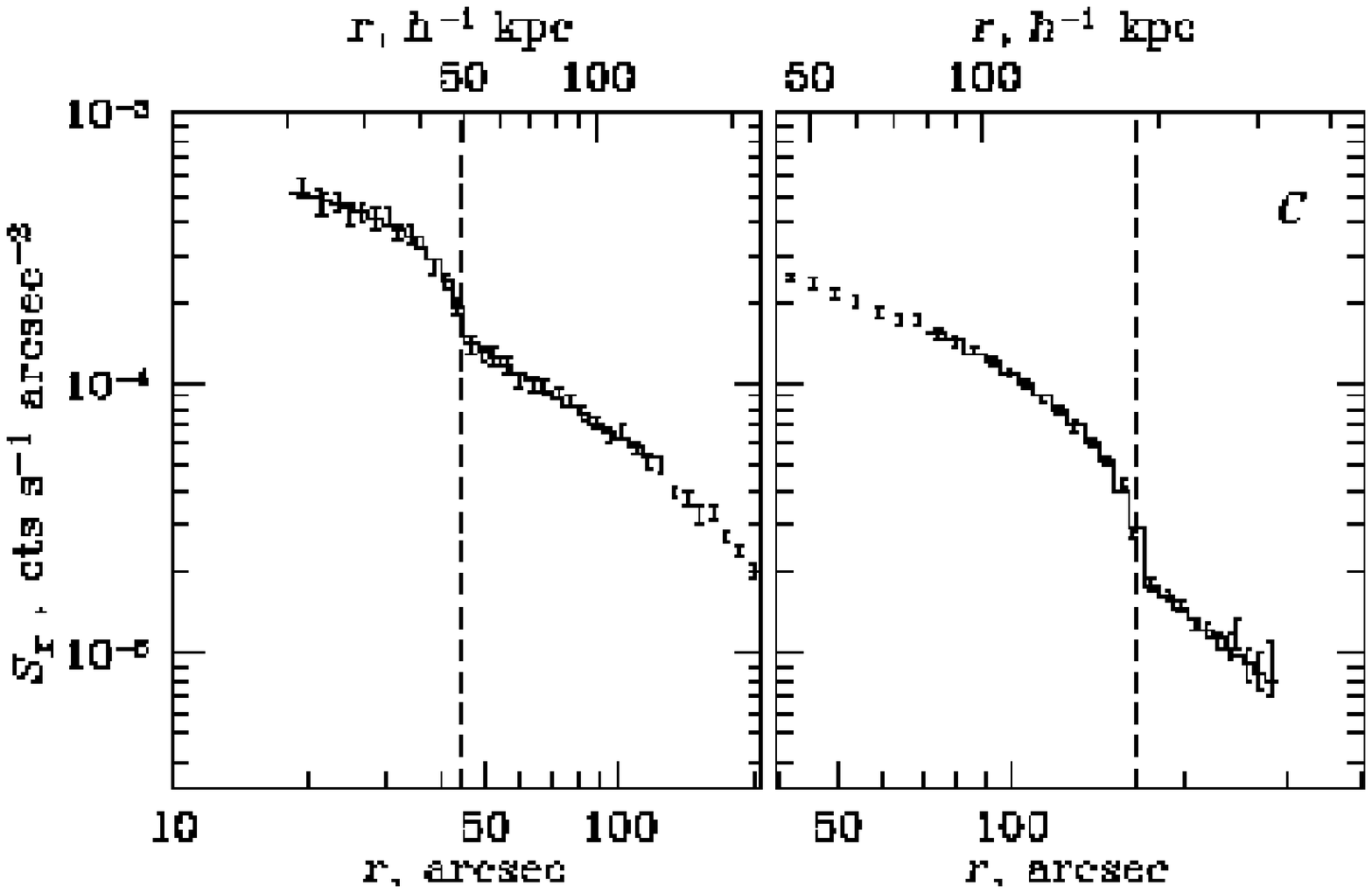}{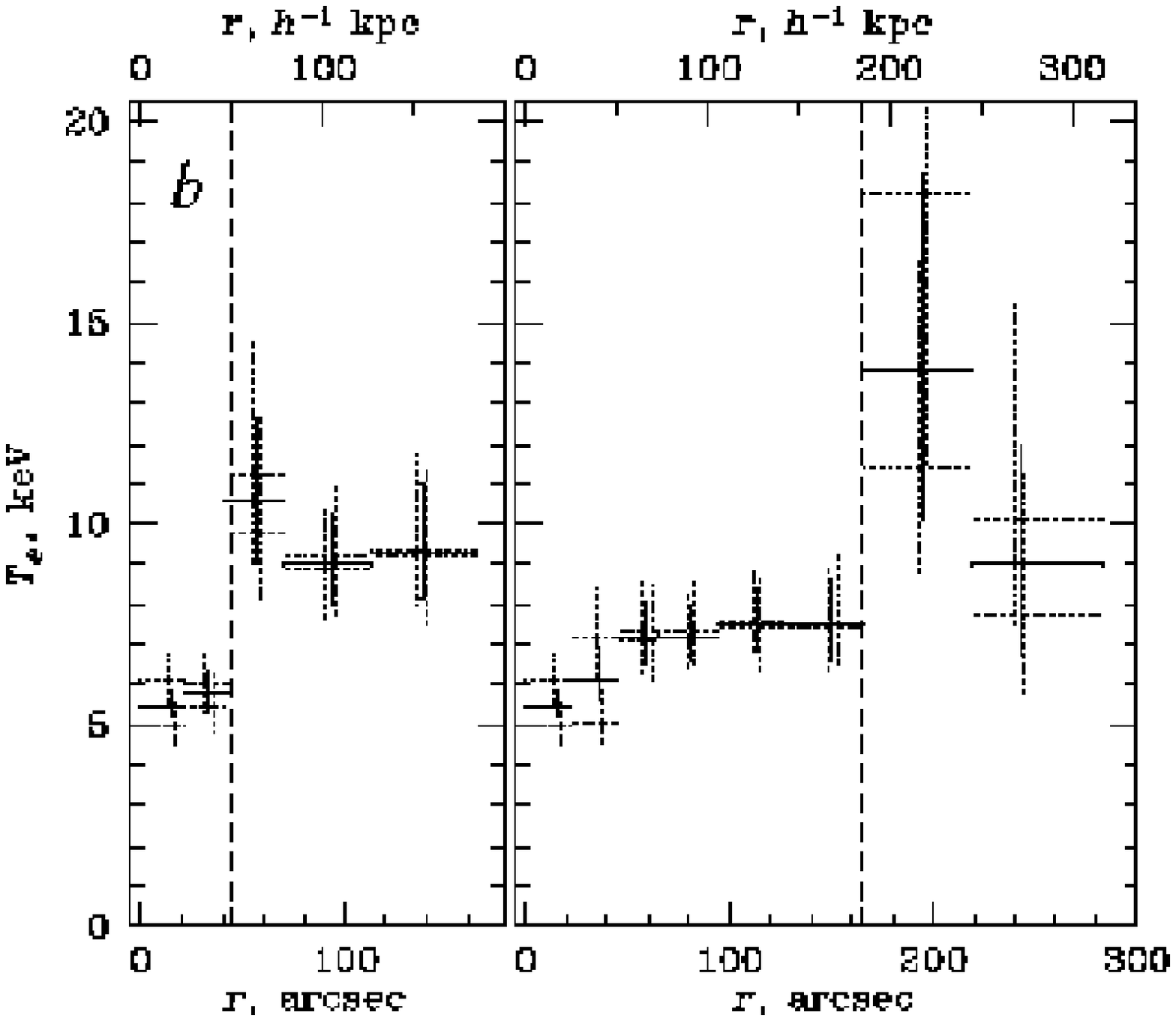}
\caption{
% (Left) X-ray image of A2142 with overlaid the contours of the considered
% sectors where breaks in surface brightness were detected.
(Left) Surface brightness profile across the two edges in A2142.
(Right) Temperature profiles along the same sectors.
(From Markevitch et al. 2000).
} \label{cond} \end{figure}

The heat stored in the intracluster plasma is conducted down any 
temperature gradient present in the gas in a way that can be described
through the following equations (Spitzer 1962, Sarazin 1988):
\begin{equation}
q =  \kappa \ \frac{d (kT_{\rm e})}{d r},
\end{equation}
where $q$ is the heat flux, $T_{\rm e}$ is the electron
temperature, and $\kappa$ is the thermal conductivity
that can be expressed in term of the density, $n_{\rm e}$,
the electron mass, $m_{\rm e}$, and the electron mean free path,
$\lambda_{\rm e} \approx 30.2 \left(\frac{kT_{\rm e}}
{\mbox{10 keV}}\right)^2 \left(\frac{n_{\rm e}}{\mbox{10$^{-3}$ cm$^{-3}$}}
\right)^{-1}$ kpc, as (Cowie and McKee 1977)
$ \kappa = 1.31 \ n_{\rm e} \ \lambda_{\rm e} \ \left(
kT_{\rm e}/ m_{\rm e} \right)^{1/2}$ 
$ = 8.2 \times 10^{20} \ \left( kT_{\rm e} / \mbox{10 keV} \right)^{5/2}$ 
erg s$^{-1}$ cm$^{-1}$ keV$^{-1}$. 

The maximum heat flux in a plasma is given by
$q = \frac{3}{2} \  n_{\rm e} \ kT_{\rm e} \ \bar{v}$,
where $\bar{v} = d r / d \tau$ is a characteristic velocity that 
we are now able to constrain
equalizing the latter equation to eqn.~2.

From the observed plasma properties in A2142 (Fig.~2),
we note that the temperature varies by a factor of 2
on a scale length of about 10--15 kpc.
These value imply that the characteristic time, $\delta \tau$, required to
erase the electron temperature gradient and due to the action of the
thermal conduction alone would be few times $10^6$ years.
When this time interval is compared with either the core 
crossing time of the interacting clumps of about 10$^9$ yrs
or a dynamical timescale of about $>2 \times 10^7$ yrs
from the model suggested by Markevitch et al.,
we conclude that thermal conduction needs to be suppressed by 
a factor larger than 10 and with a minimum characteristic 
value enclosed between 250 and 2500.
The frequency of the occurance of similar structures
in other cluster cores will be important in establishing
the timescale for their formation and duration and, hence, 
improve the constraint on the thermal conduction in the
intracluster plasma.

This result is a direct measurement of a physical
process in the ICM and implies that thermal conduction
is particularly inefficient within 280 $h_{70}^{-1}$ kpc of the
central core. The observed sharp temperature boundaries
also mean that mixing and diffusion are minimal. 
The gas in the central regions of many clusters has a
cooling time lower than the overall age of the system, so that a slow
flow of hotter plasma moves here from the outer parts to maintain
hydrostatic equilibrium. In such a cooling flow (e.g. Fabian 1994),
several phases of the gas (i.e. with
different temperatures and densities) are in equilibrium and would
thermalize if the conduction time were short. The large suppression of
plasma conductivity in the cluster core allows an inhomogeneous,
multi--phase cooling flow to form and be maintained, as is found from
spatial and spectral X-ray analyses of many clusters (e.g. Allen et
al. 2001).

\section{Interaction of radio source and cooling flow in the 
Perseus cluster}

The Perseus cluster (Abell~426) at a redshift of 0.0183 is the
brightest cluster in the X-ray sky.
A radio source, 3C84, is powered at the centre of the
central galaxy, NGC1275 (Fig.~3).
The \chandra X-ray observatory observed its core
for about 24 ksec (Fabian et al. 2000).
X-ray colours can be obtained to reconstruct the temperature
and absorption maps. The temperature map (Fig.~3)
shows that the gas is cooler towards the centre and, in particular, 
the coolest one lies along the rims of the inner radio lobes. 
From the surface brightness
we can obtain the gas density and, thus, radiative
cooling time that appears shortest in the very central regions.
There is no evidence for heating due to the radio source beyond
the lobes, that are probably cleared of cooler gas.
No sign of strong shocks is present.

\begin{figure}[hbt]
\plottwo{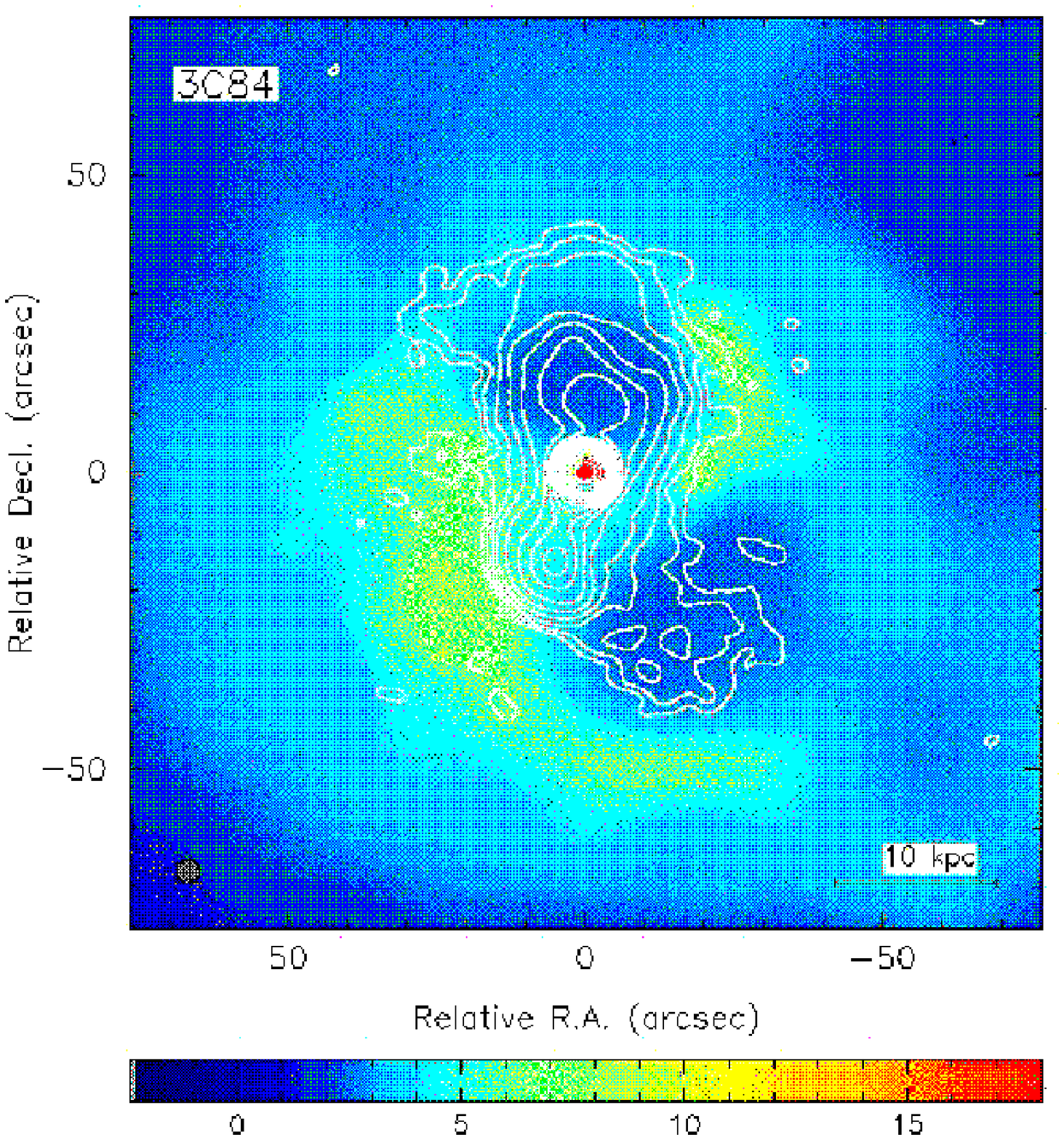}{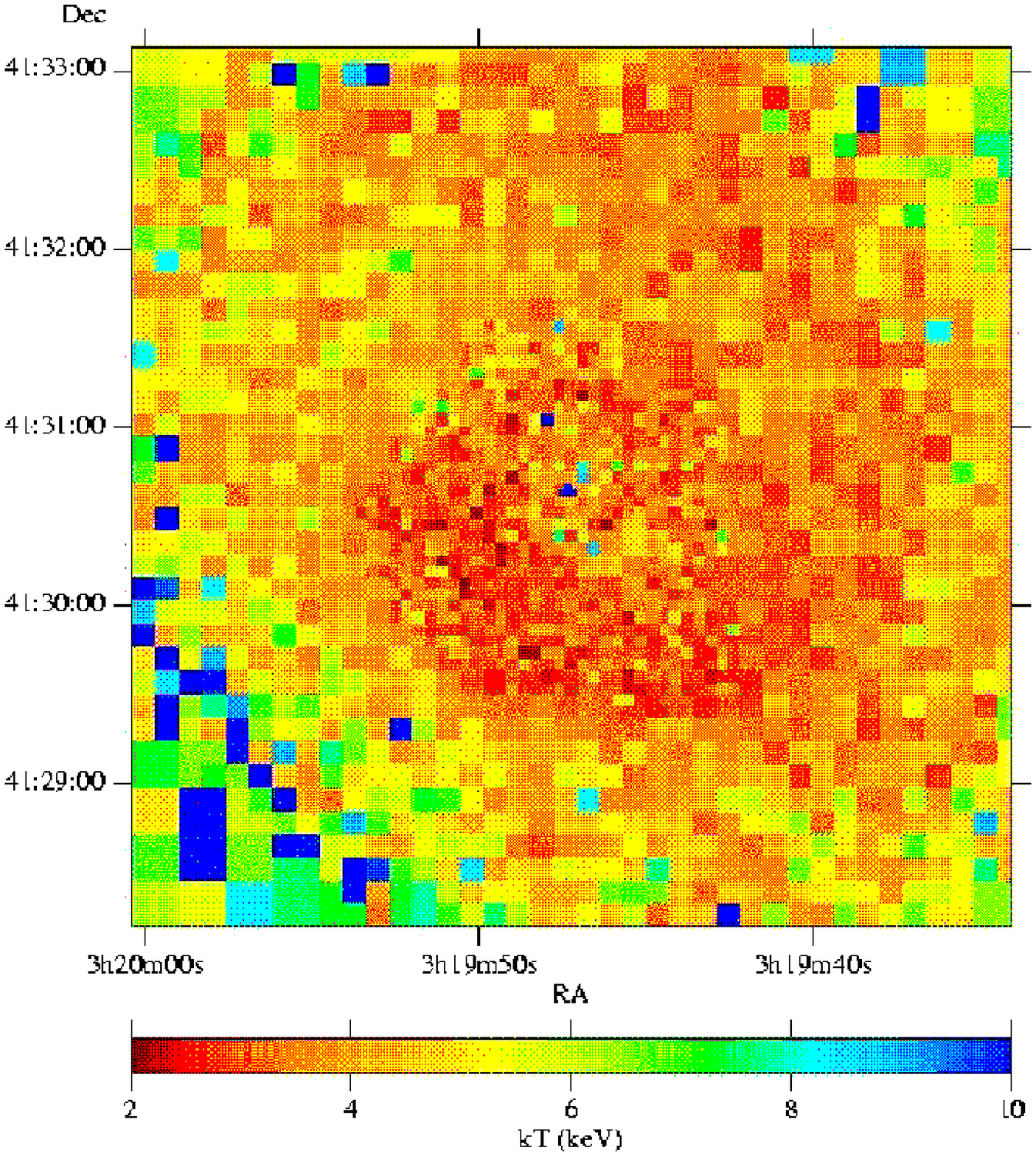}
\caption{
(Left) Radio image (1.4 Ghz restored with a 5 arcsec beam; produced
by G. Taylor) overlaid on adaptively smoothed 0.5--7 keV X-ray map.
(Right) Temperature map of the core of the Perseus cluster
obtained from X-ray color ratios (Fabian et al. 2000).
Note that the coolest gas ($T \sim$ 2.5 keV) with a cooling time
of about 0.3 Gyr lies in the rim around the lobes to the North
and East.
} \label{pers} \end{figure}

\section{Dynamical state of the core of A1795}

Abell 1795 is a nearby ($z=0.063$) rich cD galaxy cluster 
well studied at optical, radio and X-ray wavelenghts.
It has been observed for 18.4+19.4 ksec from \chandra
(Fabian et al. 2001b, Ettori et al. 2001; Fig.~4).

\begin{figure}[hbt]
\plottwo{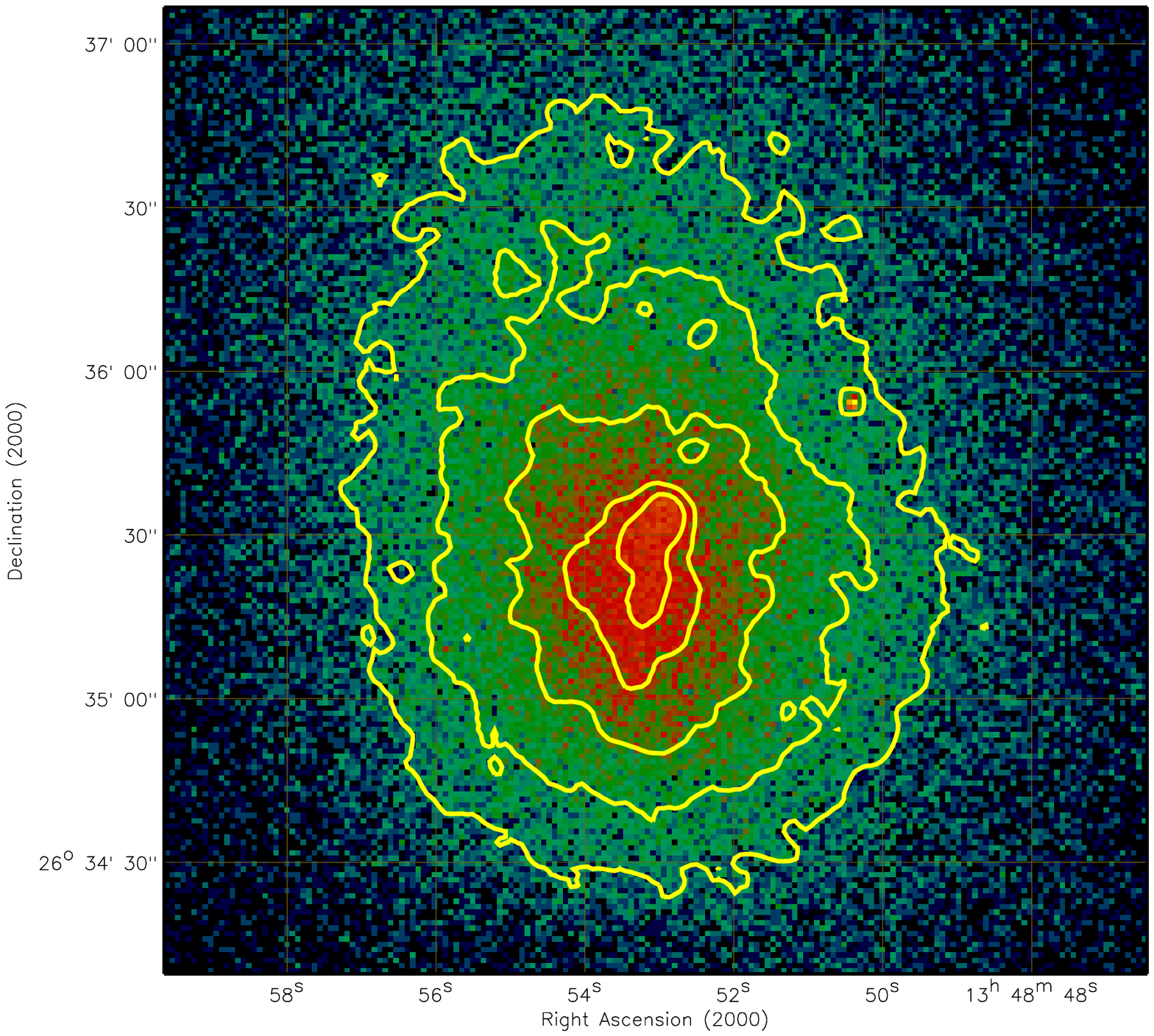}{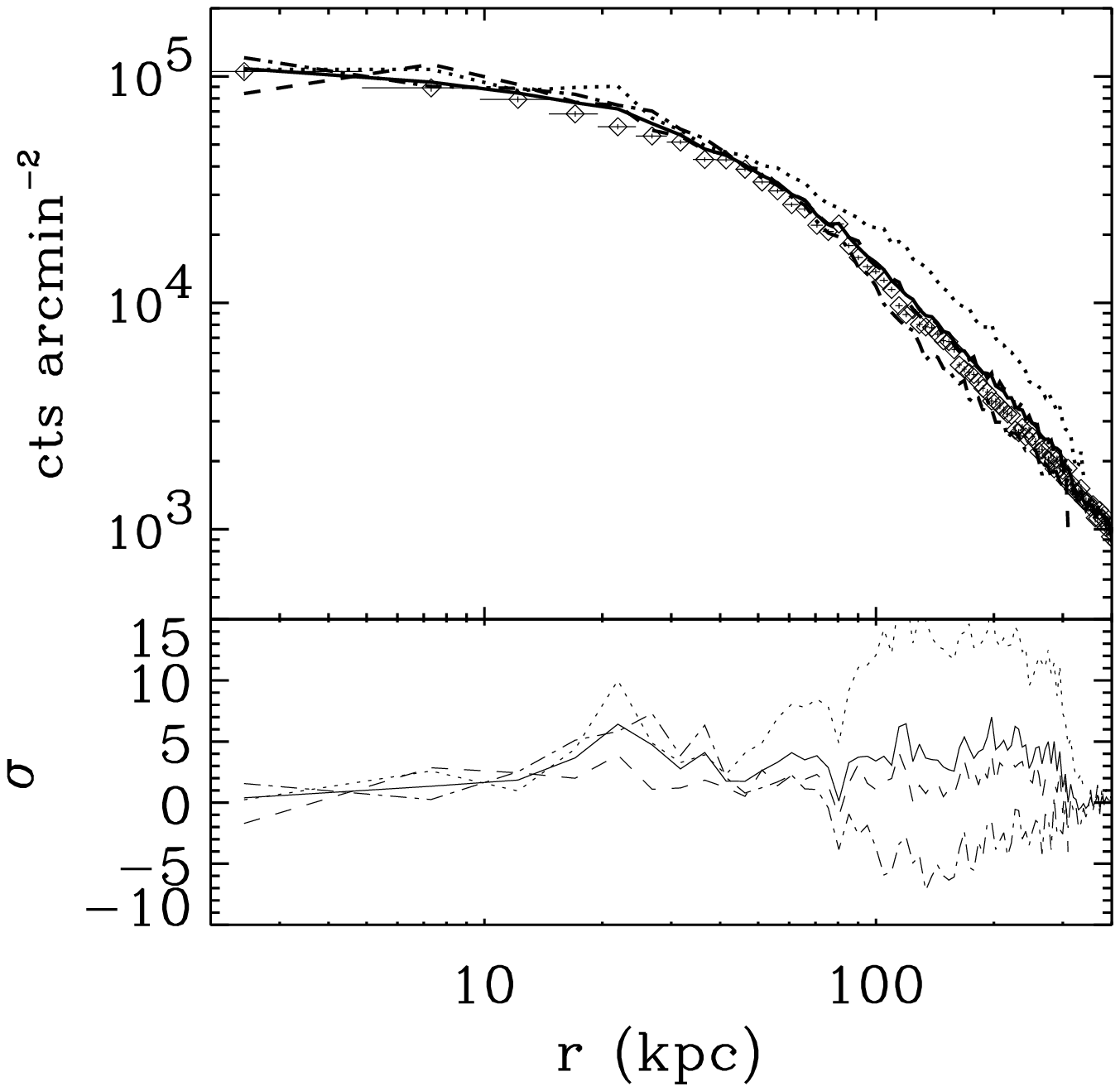}
\caption{
(Left) Smoothed image of the \chandra observation of A1795.
In the centre, it is evident an elongated feature with the cD galaxy
corresponding to the bump in brightness to the North
(Fabian et al. 2001b).
(Right) Brightness profiles extracted from four sectors ($x$ axis at $0\deg$,
$y$ axis at $90\deg$):
$50\deg-135\deg$ (``Northern excess", dotted line),
$135\deg-170\deg$ (dashed line), $170\deg-270\deg$ (``Southern deficit",
dot-dashed line), $270\deg-50\deg$ (diamonds).
The azimuthally averaged profile is represented with a solid line.
The residuals are plotted with respect to the ``$270\deg-50\deg$" profile.
Note that, starting at about 50 kpc, the excess in the North-East and the deficit
in the South-East become evident with deviations respective to 
the West region of about $10 \sigma$.
} \label{a1795} \end{figure}

To constrain the physical quantities of the ICM, 
we apply the deprojection analysis, assuming that the X-ray is
spherically symmetric and the plasma is in hydrostatic 
equilibrium with the underlying gravitational potential (cf. Fig.~1).
In our spectral analysis, we are able to resolve in 5 radial bins 
the gas temperature profile in the inner 100 kpc radius with a relative
uncertainty less than 5 (10) per cent for the projected (deprojected)
temperature. The electron density is estimated from the spectral best-fit
results. Then, we apply directly the equation of hydrostatic
equilibrium between the gravitational potential and the intracluster
plasma to estimate the mass profile, $M_{\rm X}(<r)$ (Fig.~5).
   
\begin{figure}[hbt]
\plottwo{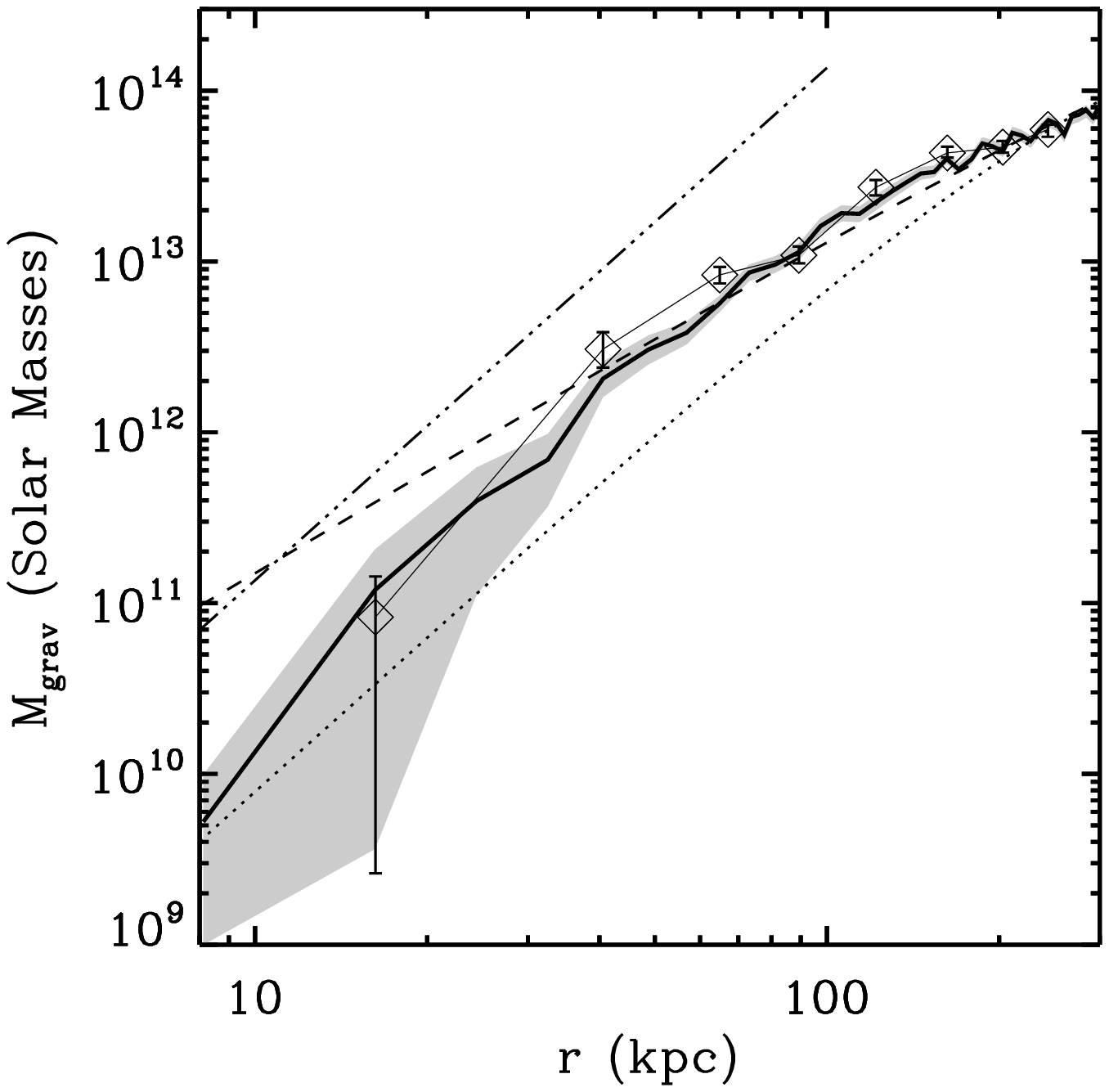}{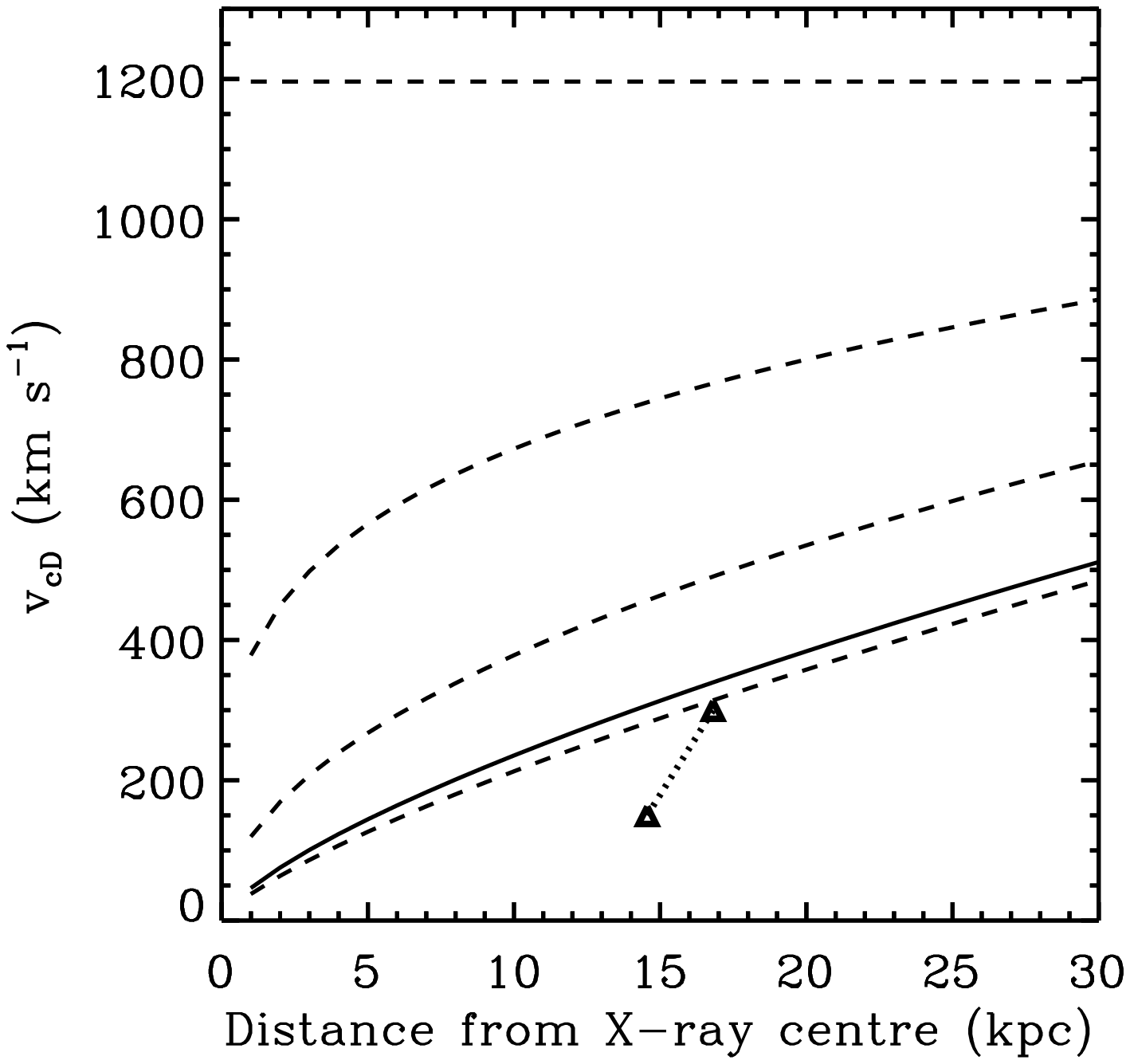}
\caption{(Left panel) Gravitational mass profiles.
The solid line is the mass profile obtained through
the hydrostatic equilibrium equation applied to the results
from the spatial deprojection on the gas temperature and density 
profiles.
The {\it diamonds} are the total mass values obtained
using the deprojected spectral temperature and density measures.
These values are compared with the best-fit results from \pspc data
(Ettori \& Fabian 1999) using a $\beta-$model (dotted line) and a 
gas density model in hydrostatic equilibrium with a Navarro, 
Frenk, \& White (NFW, 1997) potential (dashed line).
The three-dots-dash line indicates the upper limit from an assumed
tidal shear in the $H_{\alpha}$ filament due to the central 
cluster potential.
(Right panel) 
Predicted velocity of the cD galaxy under the effect of different
gravitational potentials as a function of the separation from the
X-ray centre assumed to be consistent with the deepest point of the
potential well. The solid line indicates the estimated potential and the
dashed lines are for dark matter density profiles with power law index
of --0.5, --1 (as in NFW), --1.5 and --2 upwards, respectively. 
The dotted line shows the range of the values in the velocity--separation 
space for angles with respect the line-of-sight
between 0 and 60 degrees (at $0\deg$, these values are 150 km s$^{-1}$ and
about 9 arcsec, respectively).
} \label{fig:dm}
\end{figure}

In particular, we can investigate the slope of a power law expression
of the dark matter profile, $\rho_{\rm grav} = \rho_0 (r/r_0)^{-\alpha}$,
that is a proper approximation for the inner part of the cluster potential,
integrating it over the volume ($M_{\rm grav}(<R) = 
\int_0^R 4\pi \rho_{\rm grav} r^2 dr \propto R^{3-\alpha}$)
and fitting the obtained mass profile to the one observed
(Fig.~5) between 10 and 100 kpc, where
the discrepancy among the different models is more significant.
We measure $\alpha = 0.59^{+0.12}_{-0.17}$ 
(1 $\sigma$; in the range 0.27--0.81 at the
90 per cent confidence level), 
that indicates a remarkable
flat profile suggesting the presence of a core.

The surface brightness distribution (Fig.~4)
suggests that the intracluster medium has an unrelaxed nature that
might introduce a kinetic pressure component in the hydrostatic
equilibrium equation so raising the total effective mass in the
central core. This could possibly explain the flattening of the
gravitating mass profile. Assuming that underlying potential is
described from the potential in Navarro et al. (1997) 
and $\Delta M$ is the difference with respect to
the observed mass estimates, a bulk motion with velocity (assumed
independent from radial position) of $\sim$ 300 km s$^{-1}$ is
required from the equation $d (v^2 \rho_{\rm gas})/dr = - (G \rho_{\rm
gas} \Delta M) / r^2$. This value is lower than the gas sound speed of
about 800 km s$^{-1}$ and definitely subsonic.
The shape of the gravitational potential is also
consistent with the detected motion of the cD galaxy. Oegerle \& Hill
(1994) measure a cD velocity redshifted relative to the cluster of
$\sim$ 150 km s$^{-1}$. The emission-line velocity map in Hu, Cowie, \& Wang
(1985) shows that the H$\alpha$/X-ray filament has the same velocity
of the cluster and is blueshifted from the cD velocity by still 150 km
s$^{-1}$. In Fig.~5, the cD velocity is estimated as
$v_{\rm cD} = (2 G M_{\rm grav} / r)^{0.5}$, where $G$ is the
gravitational constant and $M_{\rm grav}$ is the total gravitating
mass within the radius $r$.  Fixing
the gravitational mass at $r$= 100 kpc, we can change the power law
index $\alpha$ and investigate the behaviour of the cD velocity. 
A cD velocity larger than 400 km
s$^{-1}$ is required for potential well described by a dark matter
density profile steeper than $r^{-1}$. Given the observed velocity
offset along the line of sight, the presence of a density profile flatter 
than a NFW profile is then required.

\subsection{Cooling gas in A1795}
The gas in the core of A1795 would cool radiatively in about $10^9$ years 
(3.8 $\times 10^8$ yrs in the central 10 kpc radius, with 10th and 90th
percentile of 3.2 and 4.4 $\times 10^8$ yrs, respectively),
approaching the estimated age of the Universe at the cluster redshift
of $1.2 \times 10^{10}$ years at about 200 kpc (Fig.~6).
This can be considered as an upper limit on the region where
strong cooling is taking place.
In particular, the physical extension of the central region
where a X-ray soft component dominates the emission can be obtained 
from the flattening at about 100 kpc of the X-ray colour profile 
shown in Fig.~6. 

The deviation between the spatial and spectral results on the 
$\dot{M}$ profile, apart from the central 2 bins where the assumed
spherical geometry is probably inappropriate given the observed
cooling wake, is due to limitations on the validity of the steady--state 
cooling flow model.
The dynamical scenario, in which cooling flows establish 
and evolve, considers merging with infalling substructures 
that interrupt the subsonic flow of material.

\begin{figure}
\plottwo{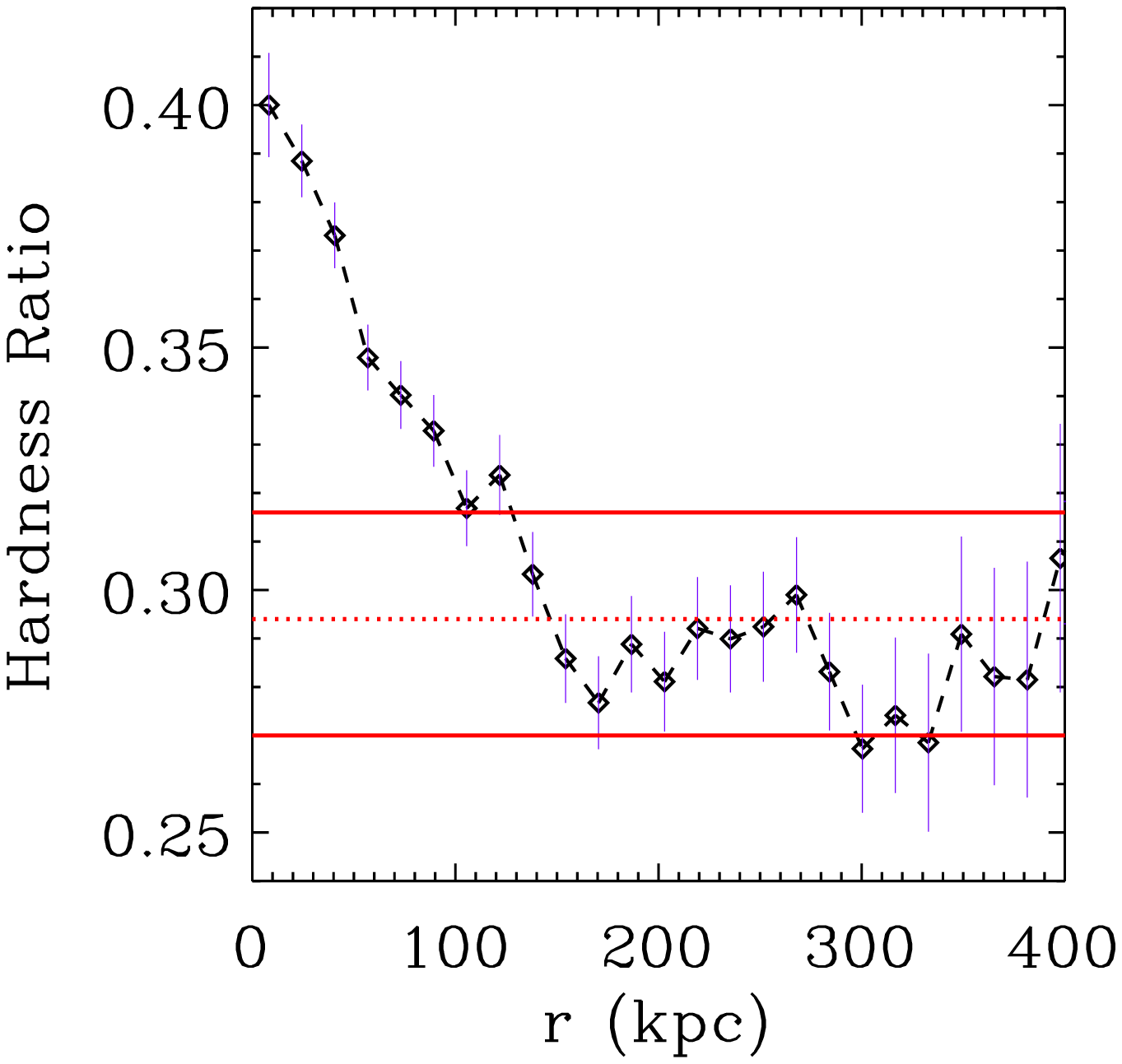}{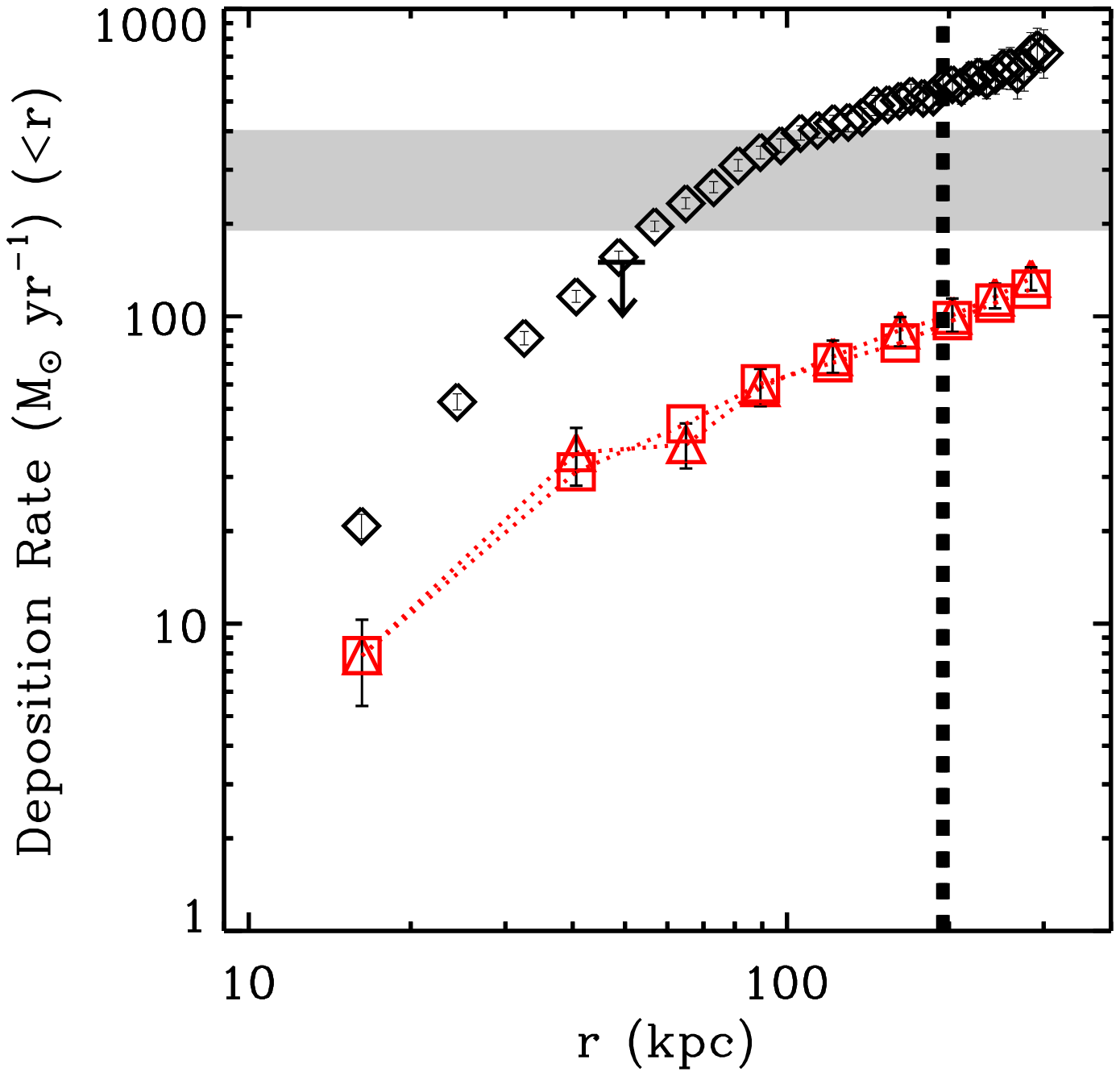}
\caption{
(Left) Ratio between counts with energy $<1.5$ and $>1.5$ keV.
(Right) Spatial ({\it diamonds}) and spectral 
({\it triangles}: from concentric rings; {\it squares}: from annuli)
deprojection results on the integrated
deposition rate, $\dot{M} (<r)$. The RGS-XMM upper limit is indicated by the downward arrow. The vertical dashed line corresponds 
to the radius of 191 kpc where $t_{\rm cool} = t_{{\rm H}_0}$.
The dashed region shows the 90\% confidence level from 
\asca analysis (Allen et al. 2001).
Note that a break at $\sim$100 kpc is present in both the two profiles
and in the colour ratio profile.
}
\label{fig:mdot}
\end{figure}

Recent analysis of \xmm data of A1795 (Tamura et al. 2001) does
not show detectable emission from gas cooling below 1--2 keV. Our
lower limit from the deprojection of the best-fit gas temperature in
the central 20 kpc radius is 1.8 keV at 90 per cent confidence level.
When an isobaric cooling flow component is considered to model the \xmm
Reflection Grating Spectrometers (RGS) spectra of the central 30\arcsec
radius region, an upper limit (90\% level of confidence) of 150
$M_{\odot}$ yr$^{-1}$ is obtained (Tamura et al. 2001). This value is
larger than our value of 43 $M_{\odot}$ yr$^{-1}$ (90\%
limit of 56 $M_{\odot}$ yr$^{-1}$) measured within the same
region of 30 arcsec from the X-ray centre. 
The \chandra-determined cooling flow therefore appears to be consistent 
with the present \xmm RGS constraint. Within the central 200 kpc, 
we estimate $\dot{M}$ of about 100 $M_{\odot}$ yr$^{-1}$ ($<$ 121 $M_{\odot}$ 
yr$^{-1}$ at 90\% c.l.).
In the model that we adopt, the cool emission is suppressed with 
an intrinsic absorption of about 3 $\times 10^{21}$ particle cm$^{-2}$.
If we use the same model that describes the RGS spectra, fixing
the outer thermal component to 6.4 keV and the Galactic column density 
to 3 $\times 10^{20}$ cm$^{-2}$, and not including any intrinsic absorption,
we measure a normalization for the cooling flow component
of 85 $M_{\odot}$ yr$^{-1}$ (90\% confidence limit range
of 76--92 $M_{\odot}$ yr$^{-1}$) but we obtain a considerably worst $\chi^2$.
The intrinsic absorption can be interpreted as one method to suppress
the line emission from gas below $\sim$ 1.5 keV. Other methods 
and related issues are discussed by Peterson et al. (2001) and 
Fabian et al. (2001a).
More discussion on the deposition rate in A1795 is presented in 
Ettori et al. (2001).

\section{The metals distribution in the cluster cores}

The X-ray emitting plasma is about 5 times more luminous than the 
stars in galaxies and, thus, is the major reserve of both baryons and heavy
metals contained in galaxy clusters, since the Iron abundance is typically
between 0.3 and 0.5 times the solar value.
Studies on the correlation between Fe mass and light coming from 
E and S0 galaxies (Arnaud et al. 1992, Renzini 1997) conclude that 
larger amount of iron resides in the intracluster medium than 
inside galaxies and its enrichment originated through releases  
from early-type galaxies.
The processes that preferably enrich the plasma are (i) (proto)galactic 
winds (De Young 1978), that occur at early times
and are characterized from supernova (SN) type II ejecta with large abundance
of $\alpha$ elements, and (ii) ram pressure stripping (Gunn \& Gott 1972), 
that takes place on longer time scales due to the continuous accretion of 
fields galaxies in the cluster potential well and produces mostly 
SN Ia ejecta.
What discriminates between these two main processes is the different
elemental mass yields: SN Ia ejecta tend to be rich in Ni, whereas
SN II ejecta present larger ratio between $\alpha$ elements 
(e.g., O, Mg, Ar, Ca, S, Si) and Fe.
Recent evidence of iron gradients in cluster cores with decreasing 
ratio between $\alpha$ elements and Fe moving inward suggest that, while
the global intracluster metal abundances are consistent with 
SN II ejecta (Mushotzky \& Loewenstein 1997), SN Ia productions
are dominant in the central cluster regions 
(e.g Finoguenov, David \& Ponman 2000, Dupke \& Arnaud 2001).

We have investigated the metals distribution in A1795.
The radial profiles of the resolved element abundances and 
the fraction of SNIa, $f_{\rm SNIa}$, responsible for those
profiles are plotted in Fig.~7.
The fraction in number of the SNe Ia is above 60 per cent 
in the inner 200 kpc.
The amount of the total iron produced in SNIa is then 
$M_{\rm Fe, SNIa} / M_{\rm Fe} \approx$ $\left[1+ 3.5 \times (0.12/0.74)
\times (f_{\rm SNIa}^{-1}-1) \right]^{-1}$ $\approx (0.8-1)$ in the
central 200 kpc (here we assume a contribution of 0.74 $M_{\sun}$ of
iron from SN Type Ia, 0.12 $M_{\odot}$ from SN Type II and 
a conversion factor of 3.5 between numbers of Type II 
and Type Ia).
This star formation activity releases energy during
supernova explosions providing an amount of thermal 
energy per gas particle, $E \approx \eta \times 10^{51}
N_{\rm SNII} \times(\mu m_{\rm p})/M_{\rm gas}$, 
where $10^{51}$ erg is the kinetic energy released by one SN, 
$\eta \ (\approx 0.1)$ is the efficiency of this kinetic energy 
in heating the ICM through galactic winds. 
When typical values for the inner part of A1795 
are adopted, $E \approx 0.2-0.6$ keV, i.e. only about 8 per cent of 
the thermal energy per particle measured (cf. Fig.~1).

\begin{figure}
\plottwo{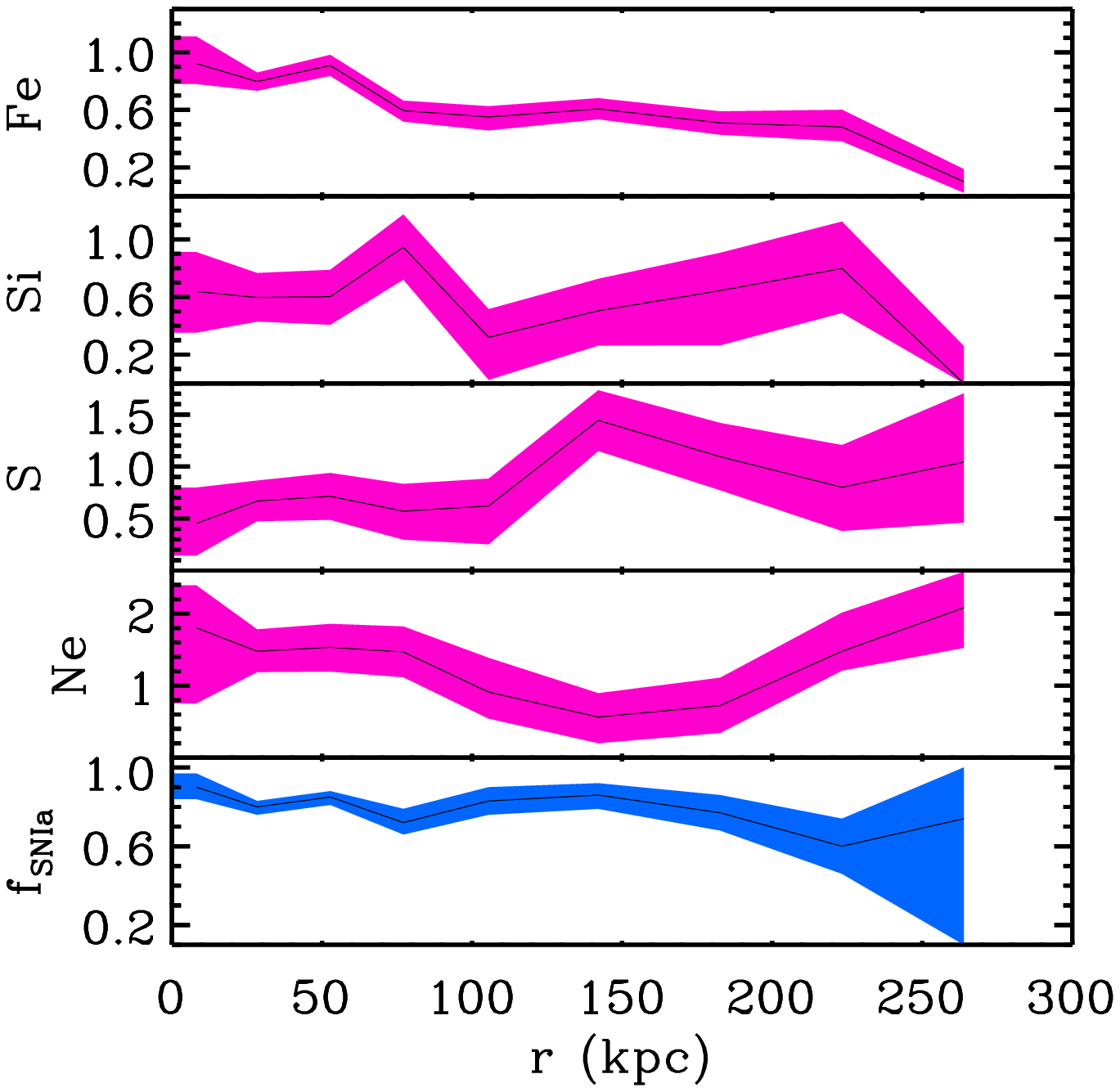}{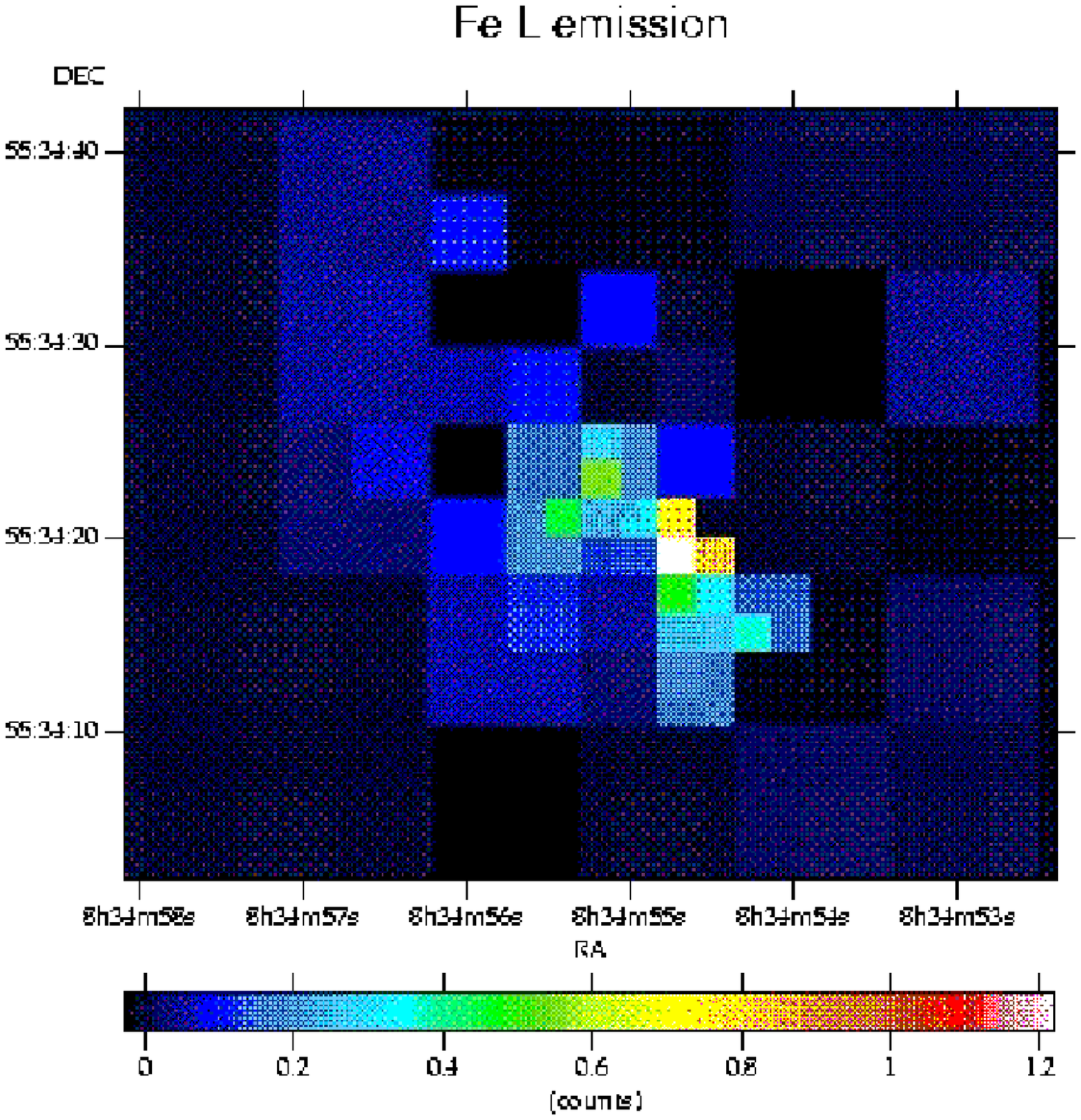}
\caption{(Left) A1795 (Ettori et al. 2001): 
abundance respect to the solar value of Fe, Si, S, Ne
as in Grevesse \& Sauval (1998).
The bottom panel provides the fraction of SNIa that are aspected to contribute
to the enrichment of the intracluster medium.
(Right) 4C+55.16 (Iwasawa et al. 2001):
adaptively pixel-binned \chandra
image of the continuum-subtracted Iron-L emission
of the core of the cluster at redshift 0.24. The plume 
located at 3 arcsec to the South-West of the centre
has a spectrum that requires a metal abundance larger than 2 solar
(90\% confidence level) and an abundance pattern typical of 
SuperNovae Ia.
} \label{fig:met}
\end{figure}

\section{Conclusions}

The \chandra observatory is providing measurements on 
arcsec scale (tens of kpc for nearby objects) of the plasma 
temperature and metals abundance with relative 
uncertainties of about 10 per cent.

The gas temperature profile raises with radius up to the 
cooling radius and then remains constant and equal
to the virial value, $T_{\rm vir}$, within few hundred 
kpc. Therefore, two temperatures, at least, are individuated 
in the cluster plasma, $T_{\rm vir}$ and $\sim T_{\rm vir}/3$
in the very central region.
Sharp jumps in the temperature/density distributions are
seen in many cluster cores, suggesting that the thermal conduction
is suppressed and the ICM is inhomogeneous.

There is not evidence for gas cooling below 1--2 keV. 
If strong shocks in cores are responsible for heating the 
gas and stopping and/or reducing the cooling flow, then
almost 10 per cent of the observed cooling flow clusters
should show evidence of this activity (cf. David et al. 2001).
No detection of strong shocks has been obtained up to now
(cf. also the Perseus case).
Peterson et al. (2001) and Fabian et al. (2001)
discuss more in depth the problems concerning
to the cooling flow scenario after \xmm and \chandra
observations.

Finally, the cores of cooling flow clusters show hints of
a positive gradient within 20--40 kpc 
(see Glenn Morris's contribution in this volume)
and negative outward
in the azimuthally averaged radial distribution of the metals.
However, strong azimuthal variations are observed
(e.g. in the core of the cluster around 4C+55.16, 
Iwasawa et al. 2001; Fig.~7).
SuperNovae type Ia appear to be dominant in the enrichment
process of these regions.

\end{document}